\documentclass[a4paper]{elsarticle}
\pdfoutput=1 

\usepackage{lineno}
\usepackage[utf8]{inputenc} 
\usepackage{amssymb}

\usepackage{soul}
\usepackage{xcolor}
\usepackage{textcomp}
\usepackage[version=3]{mhchem}	

\journal{Nuclear Physics A}

\begin{document}

\begin{frontmatter}

\title{\boldmath Modelling the shape of thermal pulses from low temperature detectors}


\author[a,b]{I. Nutini\corref{cor1}}
\cortext[cor1]{Corresponding author}
\ead{irene.nutini@mib.infn.it}

\author[c]{C. Bucci}

\author[b]{O. Cremonesi}

\address[a]{Dipartimento di Fisica, Universit\`{a} di Milano-Bicocca, Milano I-20126, Italy}
\address[b]{INFN -- Sezione di Milano Bicocca, Milano I-20126, Italy}
\address[c]{INFN -- Laboratori Nazionali del Gran Sasso, Assergi (L'Aquila) I-67100, Italy}

\begin{abstract}
Low temperature detectors are nowadays a technology widely used for rare events studies, such as the search for dark matter candidates and neutrino-less double beta decay. The convolution of the thermal and electrical response of these detectors results in pulses with different shapes depending on the materials, dimensions and operating conditions of the thermal sensors. It appears crucial to obtain a good description of the pulse shape, in order to possibly understand the several contributions in the thermal pulse formation. In this work, we present a general approach which allows us to model the shape of thermal pulses from different low temperature detectors.
\end{abstract}

\begin{keyword}
Low Temperature Detectors, Cryogenic Calorimeters, NTD, Pulse Shape, Signal Analysis, Rare events
\end{keyword}


\end{frontmatter}


\section{Introduction}
\label{sec:intro}
A full description of the time response of the detector is the main goal of many developers who aim at a full control of the performance through a complete model of the signal development. This is a very ambitious program which often has to cope with limited information and a detector complexity which only allow a very approximate result\cite{moseley1984,alessandrello1993model,probst1995,galeazzi2003}. 
A less demanding and ambitious goal is the identification of a set of functions which properly approximate the time development of the signals and can be fitted to the actual data through an appropriate set of parameters.
The fit procedure has a lot of practical applications, including filtering and signal discrimination, which are of crucial importance for the analysis of the detector data.

A finite combination of exponential functions defined on the complex plane is the common choice for the fit function. It provides generally satisfactory results, but unconstrained combinations suffer from convergence problems. Constraining  the parameter space and outlining possible correlations allows to get a quick and stable convergence of the fit procedure and is the main goal of the present paper.
We provide a set of basic rules for the fit of the time development of the signals from a general class of single particle low temperature detectors and we show few  applications of the method.
In section \ref{sec:ltd} we summarize some relevant information about low temperature detectors, while in section \ref{sec:thermalCirc} we introduce the equations of the electric analog circuit and we move on to the frequency domain for sampled data  which is where most applications are implemented. 
We conclude with few applications of the method from a set of representative thermal detectors in section  \ref{sec:signalTemplates}.

\section{Low temperature detectors}
\label{sec:ltd}
\begin{figure}[ht]
\centering
\includegraphics[width=0.8\textwidth]{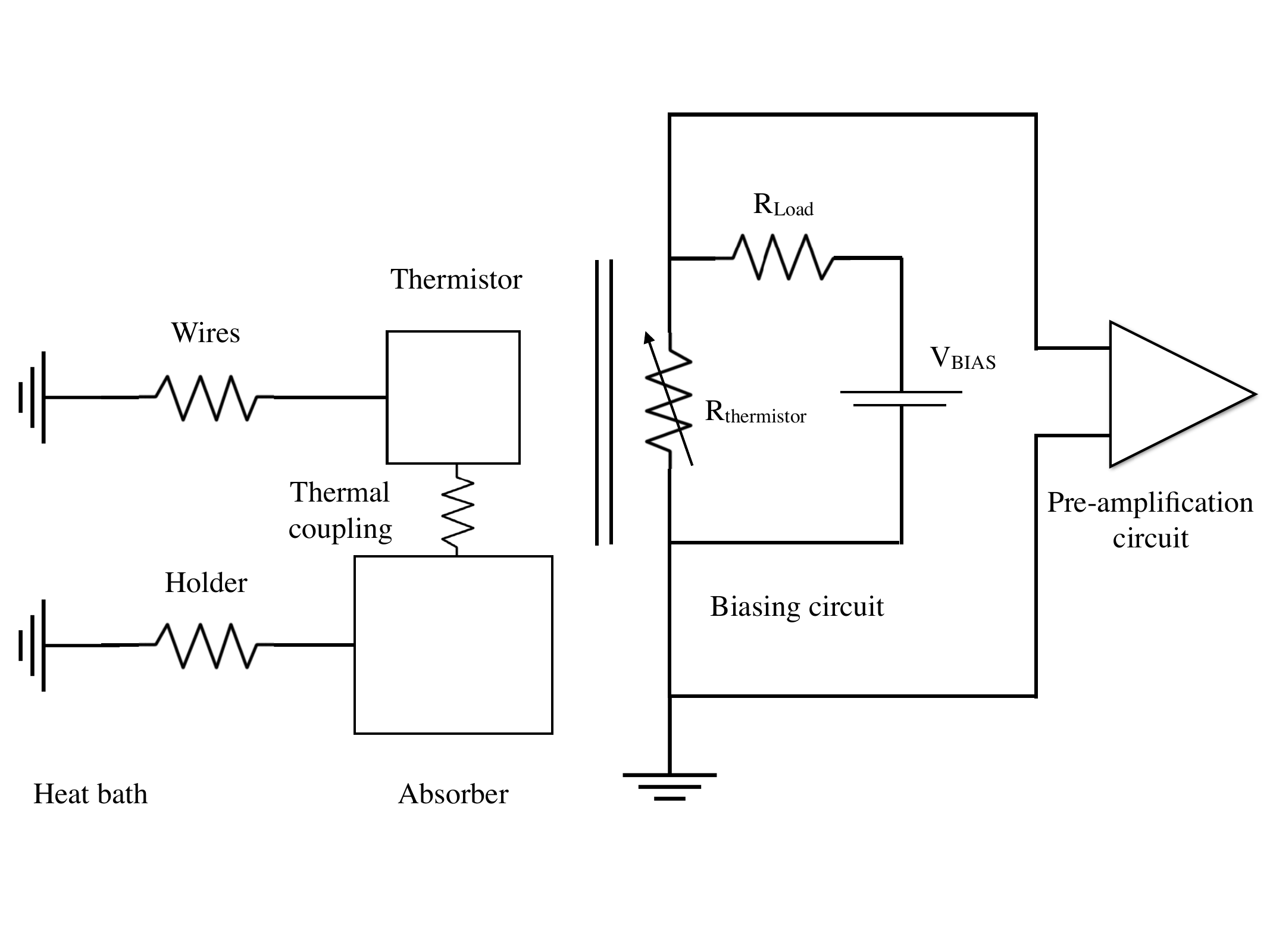}
\caption{Scheme of the detector model comprising the thermal and electric sections. This is the case of an absorber coupled to a thermistor. These two elements are operated at few mK, while the biasing and pre-amplification circuits are usually operated at room temperature.}\label{fig:ltd} 
\end{figure}

A low temperature detector (LTD) \cite{Stahl:2005} consists generally of tightly connected thermal and electrical parts (Fig.~\ref{fig:ltd})\cite{alessandrello256634,Carrettoni:2011rn}. In its simplest implementation, the thermal part describes the dynamics of the heat transfer between the different sections of the system and it is composed of a number of heat capacities interconnected by a network of appropriate thermal conductances.
Heat capacities as well as thermal conductances have generally a highly non-linear dependence on temperature so that simplifying hypotheses are needed to solve the system dynamics. The most common approximation is to assume that temperature variations are small along the dynamical evolution of the system (small signal approximation or SSA) so that thermal conductances and capacitances can be approximated by constant values.
The dependence on temperature can then be described separately by means of a parametric approach in which the behaviour of the system is analysed for different temperature distributions.
Non-linear effects (e.g. electro-thermal feedback) are not included in this picture by construction. They can be analysed in terms of dynamical properties of the  solutions and represent one of the most important features of an exact detector model.

The electric part consists of a transducer, which converts the thermal excitation to an electric measurable quantity (e.g. an electric resistance), its bias circuit and a proper filtering system (e.g. anti-aliasing filters before signal digitization). We will consider only NTD thermistors \cite{Larrabee:1984} as thermal transducers, for the discussion in this paper. 
Under proper conditions (e.g. SSA and a sufficiently wide frequency band) also the electric part can be simplified allowing an analytic description of the detector dynamics as a whole.

The result is a simple approximated model able to describe the signal development of most thermal detectors in terms of a limited number of free parameters that can be fitted to experimental data and eventually linked to the physical parameters of the system. 

\section{The electric circuit analog}
\label{sec:thermalCirc}
The electric circuit analog is the simplest way to solve the dynamics of the thermal section.
Thermal conductances and capacities are replaced by their electric analog, the temperature acts as a voltage and the thermal power as an electric current.
The resulting system consists of two coupled electric circuits whose evolution can be simply obtained exploiting the usual tools for the study of electrical systems.  

It can be easily shown that for a continuous-time linear time-invariant system, the most general form for the transfer function $H(s)$ is given by the ratio of two polynomials:
\begin{equation}
\label{eq:transfer}
H(s) = \frac{N(s)}{D(s)} = \frac {b_M s^M+b_{M-1}s^{M-1}... + b_0}{a_N s^N+a_{N-1}s^{N-1}... + a_0} = 
\frac {\prod_{j=1}^{M} (s-z_j)}{\prod_{k=1}^{N} (s-p_k)} = 
\sum_{k=1}^N \frac{r_k}{s-p_k}
\end{equation}

where M and N are the degrees of the numerator and denominator polynomials, z$_j$ and p$_k$ are known as zeroes and poles of the transfer function, and r$_k$=\,$ \lim_{s\to p_k} [(s-p_k) \cdot H(s)] $ is the  k-th pole residue. 
 
Since each independent heat capacity in the thermal section adds a pole to the transfer function, the number of poles must be greater or equal to the number of heat capacities. 

System stability then requires R\,$\equiv$\,N-M\,$\ge$\,2 which means that only systems with N\,$\ge$\,2 provide stable solutions and  zeroes are allowed only when N is strictly larger than 2.

Furthermore, it can be demonstrated that only zeroes and poles lying close to the imaginary axis dominate the dynamics of the system so that usually the circuit can be described in terms of a limited number of poles and zeroes with the region of convergence consisting of a vertical band which includes the imaginary axis and extends to the right of the pole with the largest real part.

Interesting properties of the poles can be deduced just looking at the coefficients of the denominator in eq. (\ref{eq:transfer}). For example, the Viète rule\cite{vinberg:2003} implies that\\ $\sum_{k=1}^N \, p_k$\,=\,$-a_{N-1}/a_N$, while a sufficient condition for avoiding unwanted poles with a positive real part is that all the coefficients a$_k$ have the same positive sign \cite{mead1992,curtiss1918}.

The time development of the signals is then given by the inverse Laplace transform of $H(s)$ which, by linearity, reads:

\begin{equation}
\label{eq:zepo2}
h(t) = \sum_{k=1}^{N} r_k e^{p_k t}\; ,\quad 
\sum_{k=1}^N r_k = 
\begin{cases} 
b_M / a_N , & {\rm if}\; R=1 \\ 
0, & {\rm if}\; R\geq 2
\end{cases}
\end{equation}

The most general function describing the signals of a stable low temperature detector is therefore a combination of N exponentials with coefficients which are strongly correlated, through the form of $H(s)$ (eq.\ref{eq:transfer}). 
The strength of the proposed method consists just in implementing these correlations in the expression of the fit function forcing the coefficients to fill only limited regions of the parameter space.

\begin{figure}[ht!]
\centering
\includegraphics[width=0.7\textwidth]{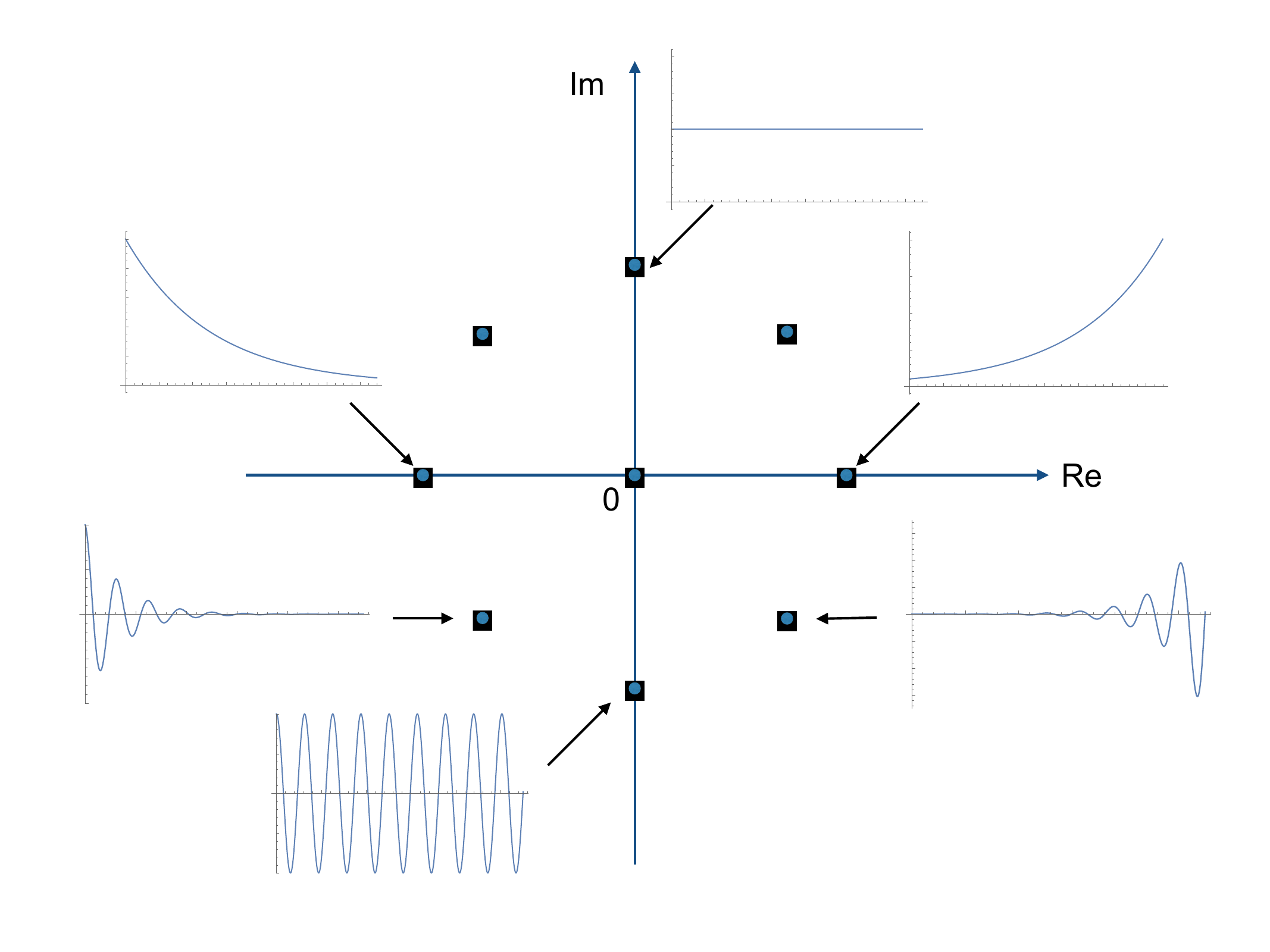}
\caption{Schematics of the time response of circuits, given the position of the poles.}\label{fig:poles} 
\end{figure}

Indeed, the pole position determines the response of the circuit, as schematically represented in Fig.\ref{fig:poles}.
In particular, poles with negative real part have exponential attenuation and complex poles generate dumped oscillations. 

It is worth to note that the thermal connection with a heat bath ensures that the LTD eventually recovers back to the initial condition so that the system is asintotically stable (R$\geq$ 2). 

Some simple rules can be helpful in identifying the role of each pole. 
Let's start with the case of real poles:
\begin{itemize}
\item the pole closer to the imaginary axis represents the longest decay time constant;
\item if no zero is present in between two adjacent real poles, then the sequence of poles alternates decay and rise time constants;
\item a zero located between two adjacent poles inverts the sequence (i.e. acts as an additional pole) and the next pole  maintains the same behaviour in terms of time constants;
\end{itemize}
Now, since for physical systems D(s) is characterized by real coefficients, the poles can only be real or pair-conjugate, the latter case being responsible of an oscillatory pattern.

Variations of the circuit parameters (capacitances, or resistances) correspond to changes of the response function or, correspondingly, of the positions of the poles which therefore draw characteristic trajectories in the complex plane as a function of the working temperature or bias parameters. 
The shapes of these trajectories  depend on the details of the adopted thermal model and are an essential ingredient of deeper studies.

So far we have discussed the continuous (or analog) case. Most applications refer however to digitized (and finite) data which require some additional consideration.

Here the transfer function is replaced by $H^d(z)$, the z-transform of the response to the unitary function which maintains a very similar form:
\begin{equation}
\label{eq:zepo}
H^d(z) = \frac {\prod_{k=1}^{M} (z-Z_k)}{\prod_{j=1}^{N} (z-P_j)}
\end{equation}
where Z$_k$ and P$_j$ are now the zeros and poles of $H^d$. 
This is not surprising since $H(s)$ and $H^d(z)$ describe the same physical system.

Now, since most of the circuit theory is built on analog systems, the relation between the continuous and discrete domains is essential in order to translate results from one domain to the other. 
The link is provided by the Fourier transforms, which represents a limiting case of the Laplace transform to the imaginary axis in the analog domain, $F(\omega)=H(s=i\omega)$, and of the z-transform to the unitary circle in the discrete domain, $F^d(\omega ) = H^d(z=e^{i\omega}$).
Furthermore, in the common case of finite discrete  sequences the  ``digital'' frequencies can only assume integer multiples of the fundamental frequency $\Omega=\frac{2\pi}{T}$, where T is the width of the time window. Thus we recover the discrete Fourier transform as a limiting case of the z-transform on a set of equidistant points on the unitary circle, $F^d(\omega^d) = H^d(z=e^{i\omega^d}$,  $\omega^d=k\cdot \Omega$, k={1,...,N}).

Now the bi-linear transformation (or warping) provides a straightforward mapping between the continuous/analog and discrete/digital frequency domains: 
\begin{equation}
    \pi\frac{f^a}{f_s}\equiv \frac{\omega^a}{2f_s}=\tan(\pi\frac{f^d}{f_s})\equiv\tan(\frac{\omega^d}{2f_s})
\end{equation}
where $f_s$ is the sampling frequency, and variables with superscript $a$ are related to analog frequencies, while the ones with $d$ superscript are the digital ones. The warping allows to directly recover the discrete Fourier transform:  $F^d(\omega^d)\,=\,F(\omega^a\to\omega^d)$.

A new strategy for the fit can therefore be devised, in the frequency domain: the discrete Fourier transform of the finite sequence of signal samples $S(\omega^d)=\mathbb{F}(s({t_k}))$ is fitted to $F^d(\omega^d)$ derived from $H(s)$ by limiting its values to the imaginary axis and then applying a bi-linear transformation. 
In particular a convenient $\chi^2$ function can be defined as
\begin{equation}
\label{eq:nufit}
\chi^2\equiv \sum_k\frac{\vert(S_k-F^d_k)\vert^2}{\mathcal{N}_k}
\end{equation}

where $\mathcal{N}$ is the noise power spectrum.

The only unknown parameters are therefore the number and location of zeroes and poles of the transfer function and experimental data can be weighted using the noise power spectrum.

\section{Signal templates for low temperature detectors}
\label{sec:signalTemplates}

The successful application of this technique relies on the knowledge of the number of dominant poles and zeroes of $H(s)$. However the detectors are usually complex and it is not straightforward to understand which are the leading contributions and which are the elements that can be neglected.
An effective way is to examine the detector pulses trying to fit them with a simple transfer function and increasing gradually the complexity whenever the results are not satisfactory.
In addition the Pad\'e-Laplace method \cite{yeramian1987analysis} can help identifying the number of time constants (i.e. the number of dominant poles)\cite{bajzer1989pade,clayden1992pade}.

In the following examples, we will report thermal pulses from different low temperature calorimetric detectors (TeO$_2$, Li$_{2}${}MoO$_4$,...) read by NTD thermistors, with the respective models for the transfer function used to fit the data.
The waveforms are acquired with 1 (or 2) kHz sampling frequency and the hardware anti-aliasing filter cutoff is larger than the signal bandwidth.

Every time a pulse is identified in the data-stream, a trigger flag is set as soon as its rising edge exceeds a definite threshold. We define as 'event' the given time interval in which the pulse is recorded; the event window length is fixed. The first part corresponds to samples acquired before the pulse, and it is characterized by a constant value B (baseline) that can be used to calculate a proxy of the temperature and to check the noise and thermal stability of the data. The position of the trigger flag is assigned with time $t_0$. Given this construction of the event window, we will consider in the detector response function for the pulse fit: a time-shift $t_0$ and a non-zero reference value for the waveform baseline ($B$). Moreover we add a pulse amplitude term $A$ multiplied to the pulse response, to take into account for different energy depositions. 

\subsection{Template pulse with two real poles}

In some cases, often regarding small calorimeters, it has been observed that the pulses are well described by only one rise and one decay constant. Given the previous considerations, the detector response can be written in terms of a transfer function $H(s)$ with 2 real negative poles. 

\begin{equation}
H(s) = \frac{1}{(s - p_1)  (s - p_2)} =  \frac{r_1}{(s - p_1)} + \frac{r_2}{(s - p_2)} \qquad \text{with} \qquad p_2 < p_1 < 0
\end{equation}
where $r_1$ and $r_2$ are the residues calculated in the two poles; given the position of the poles: $r_1 = -r_2$, $r_1 > 0$ and $r_2 < 0$.

The pulse shape in the time-domain $h(t)$ is obtained with the Laplace inverse transform of the transfer function $H(s)$, including the start time $t_0$, the baseline {\it{B}} and amplitude {\it{A}} as additional parameters. The pulse response in the time domain is then given by equation:

\begin{equation}
    h(t) = B + A \cdot [ r_1 \cdot e^{p_1 (t - t_0)} + r_2 \cdot e^{p_2 (t - t_0)} ]
\end{equation}{}

\begin{figure}[htbp]
\centering
\includegraphics[width=0.9\textwidth]{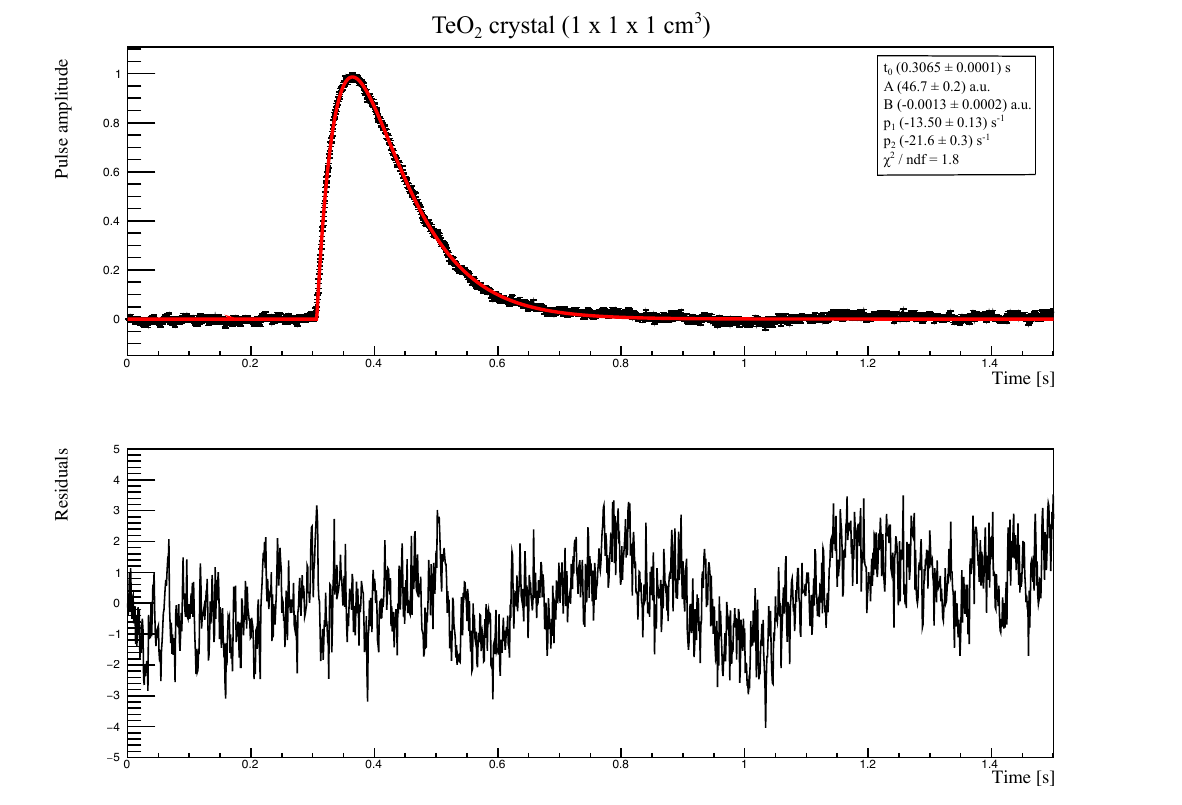}
\caption{Example: Pulse from a small TeO$_2$ calorimeter ($1 \times 1 \times 1$  $cm^3$) read by a NTD. The pulse is fit with a signal template with 2 real poles ($p_1$,$p_2$).} \label{fig:example2pRe_microTeO}
\end{figure}

This template gives a good representation of the pulse shape. Two examples are reported in the figures below. In Fig.\ref{fig:example2pRe_LD}, the pulse from a Germanium Light Detector (Ge-LD) read by an NTD \cite{Beeman:2013zva}, is reported. The Ge-LD (44 mm diameter and 175 $\mu$m thickness) is operated at $\sim$\,20\,mK, it faces the main absorber and converts the scintillation photons into phonons; the phonons in the LD are then converted into an electric signal via a NTD thermistor. In Fig.\ref{fig:example2pRe_microTeO}, it is shown the pulse from a TeO$_2$ cubic calorimeter of dimensions $1 \times 1 \times 1$  $cm^3$, coupled with an NTD, glued on a copper holder.\\ 
It is not the aim of this paper to discuss the physical meaning of the two poles describing the pulses. However we want to focus on the potential of this approach in building accurate templates for both heat and light LTD detectors.

\begin{figure}[htbp]
\centering
\includegraphics[width=0.9\textwidth]{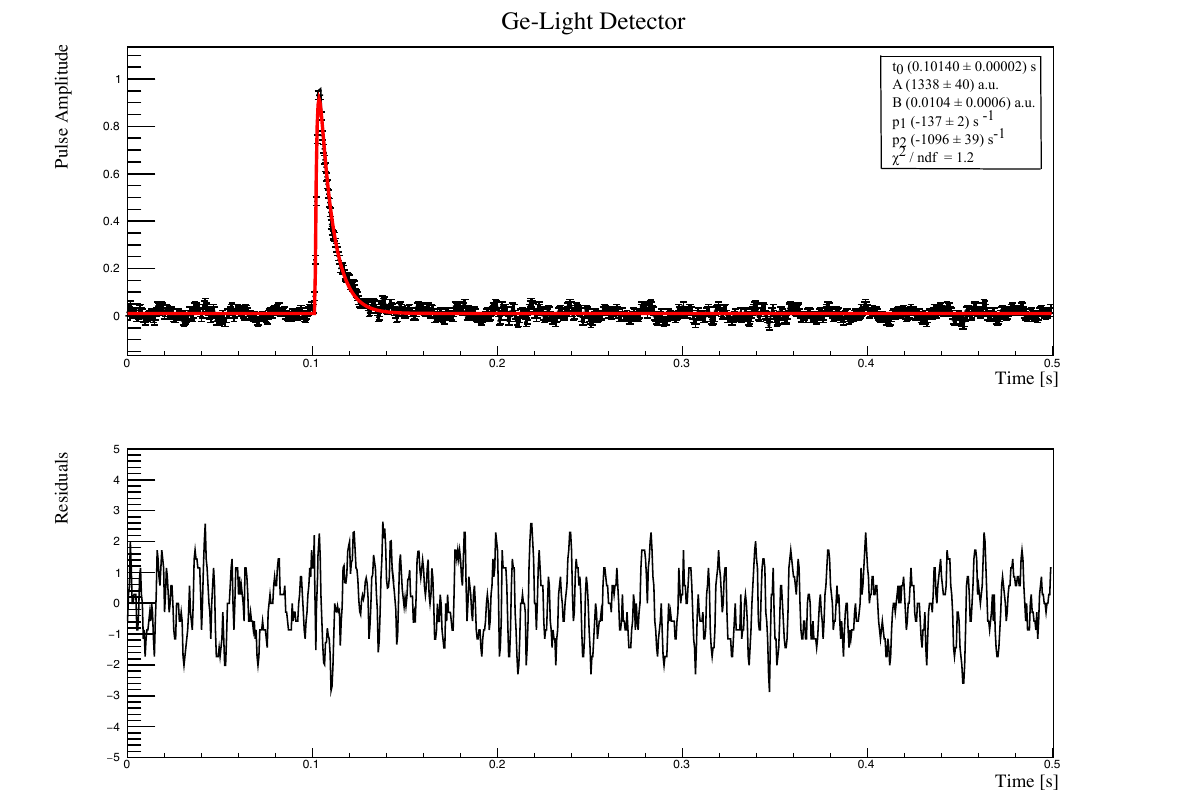}
\caption{Example: Pulse from a Ge-LD read by a NTD. The pulse is fit with a signal template with 2 real poles ($p_1$,$p_2$).} \label{fig:example2pRe_LD}
\end{figure}

\subsection{Template pulse with three real poles and one zero}
Thermal pulses from several macro calorimeters (volumes $\sim$\,tens $cm^3$, masses $\sim$\,$kg$) coupled with NTDs are often characterized by 3 dominant time constants, one rise time and two decay times \cite{Pedretti_PhD-thesis:2004,Vignati_PhD-thesis:2010,Santone_PhD-thesis:2017}. It is worth to note that the observed rise time is smaller than both the decay times.\\ 
Given these premises, the detector response cannot be described by 3 real poles only ($p_3 < p_2 < p_1 < 0$), since the rise time value, associated with $p_2$, would be intermediate between the two decay times here. It is necessary to add a zero in between the first two poles, $p_3 < p_2 < z_1 < p_1 < 0$, in order to make $p_3$ to be the rise time and ($p_1$, $p_2$) the decay times. The detector response is therefore written in terms of a transfer function $H(s)$ with 3 real negative poles and 1 zero.

\begin{equation}
H(s) = \frac{(s-z_1)}{(s - p_1)  (s - p_2) (s - p_3)}=  \frac{r_1}{(s - p_1)} + \frac{r_2}{(s - p_2)} + \frac{r_3}{(s - p_3)} 
\end{equation}

where $r_1$, $r_2$ and $r_3$ are the residues calculated in the three poles; given the position of the poles and the zero: $r_1 > 0$, $r_2 > 0$ and $r_3 < 0$.

The pulse shape in the time-domain is 
constructed as in the equation below:

\begin{equation}
    h(t) = B + A \cdot [ r_1 \cdot e^{p_1 (t - t_0)} + r_2 \cdot e^{p_2 (t - t_0)} + r_3 \cdot e^{p_3 (t - t_0)} ]
\end{equation}{}

Examples of the detector response function with 3 real negative poles and 1 zero, fitting pulses from calorimeters of different materials are reported in the figures below.

In Fig.\ref{fig:example3p1z_TeO_Q0}, it is reported the pulse from a TeO$_2$ cubic calorimeter of dimensions $5 \times 5 \times 5$  $cm^3$ read by an NTD, from the CUORE-0 experiment\cite{Alduino:2016vjd,Alfonso:2015wka}; that was fit in the time domain with the equation above. In Fig.\ref{fig:example3p1z_TeO_Q0_freq} the same CUORE-0 pulse is reported in the frequency domain, via a discrete Fourier transform; we perform the fit of its real and imaginary parts using the transfer function in the Laplace domain, converted into digitized frequencies, and minimizing the $\chi^2$ function from Eq.\,\ref{eq:chi}.  We can see from the residuals plot in Fig.\ref{fig:example3p1z_TeO_Q0_freq} [bottom] that the fit in the frequency domain is able to disentangle the noise contributions at known frequencies from the expected pulse shape for the experimental data.  Both the fits in the time domain and frequency domain show a good accuracy in reproducing the pulse shape, and have consistent results in terms of the values of the poles and the zero. 

\begin{figure}[htbp]
\centering
\includegraphics[width=0.9\textwidth]{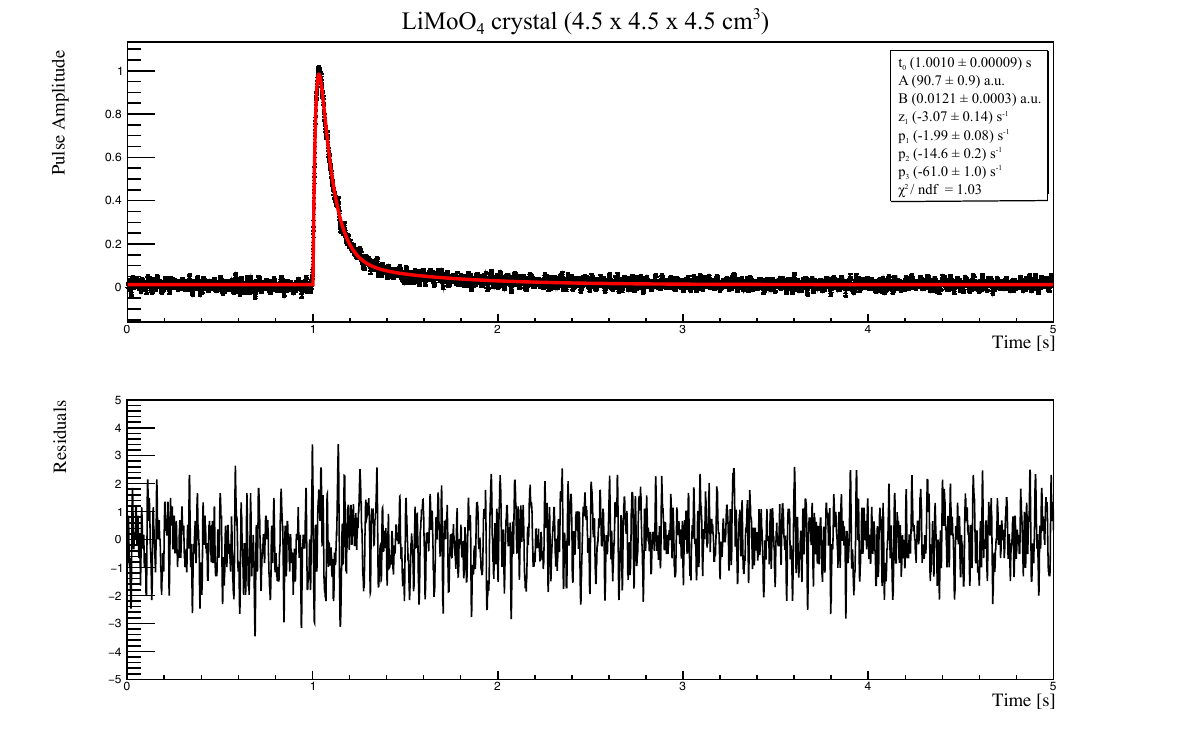}
\caption{Example: Pulse from a Li$_{2}${}MoO$_4$ macro-calorimeter read by a NTD. The pulse is fit with a signal template with 3 real poles ($p_1$,$p_2$,$p_3$) and 1 zero ($z_1$).} \label{fig:example3p1z_LMO_salaC}
\end{figure}

Li$_{2}${}MoO$_4$ crystals are also utilized as cryogenic detectors for rare events physics \cite{Cardani:2013dia,Cardani:2020hallC}; in Fig.\ref{fig:example3p1z_LMO_salaC}, there is a pulse from a Li$_{2}${}MoO$_4$ semi-cubic crystal of dimensions $4.5 \times 4.5 \times 4.5$  $cm^3$, read by an NTD, and its fit in the time domain with the transfer function with 3 real poles and 1 zero.\\

\begin{figure}[htbp]
\centering
\includegraphics[width=0.9\textwidth]{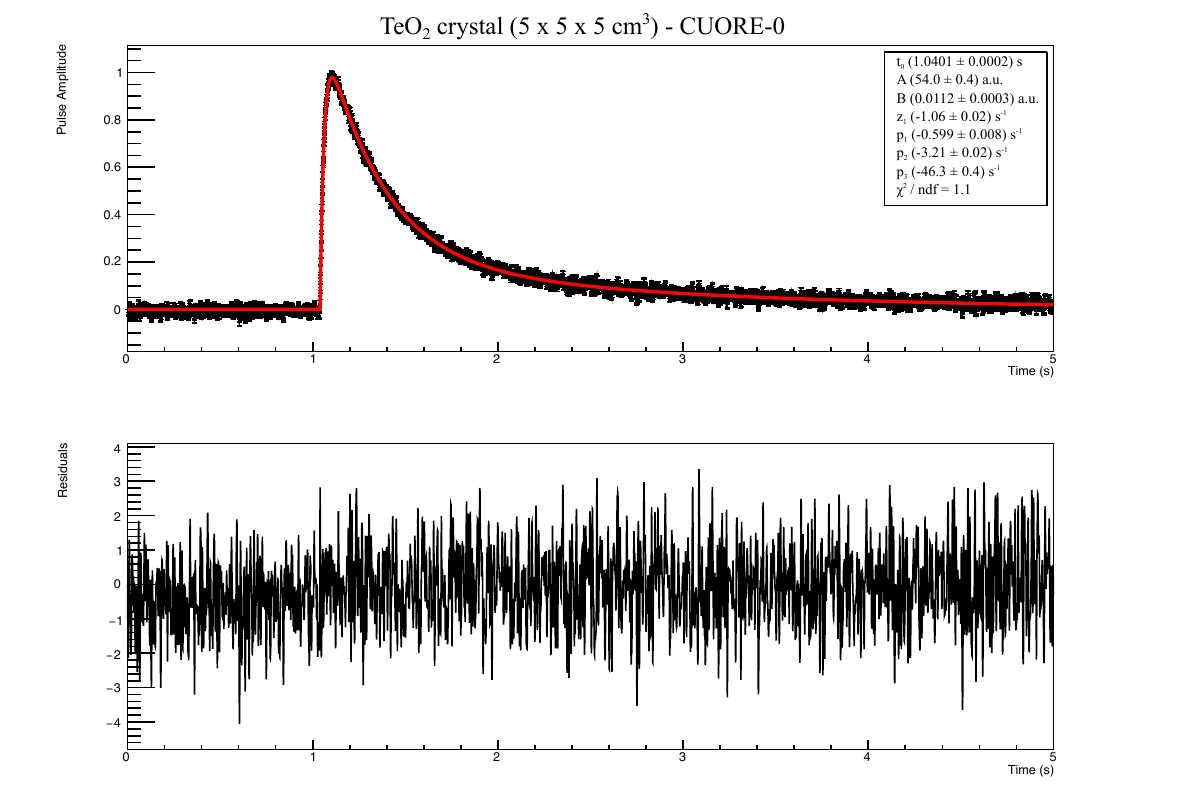}
\caption{Example: Pulse from a TeO$_2$ macro-calorimeter read by a NTD (from CUORE-0 experiment). The pulse is fit in the time domain with a signal template with 3 real poles ($p_1$,$p_2$,$p_3$) and 1 zero ($z_1$).
} \label{fig:example3p1z_TeO_Q0}
\end{figure}

\begin{figure}[htbp]
\centering
\begin{minipage}[b]{\linewidth}
 \centering
\includegraphics[scale=0.5]{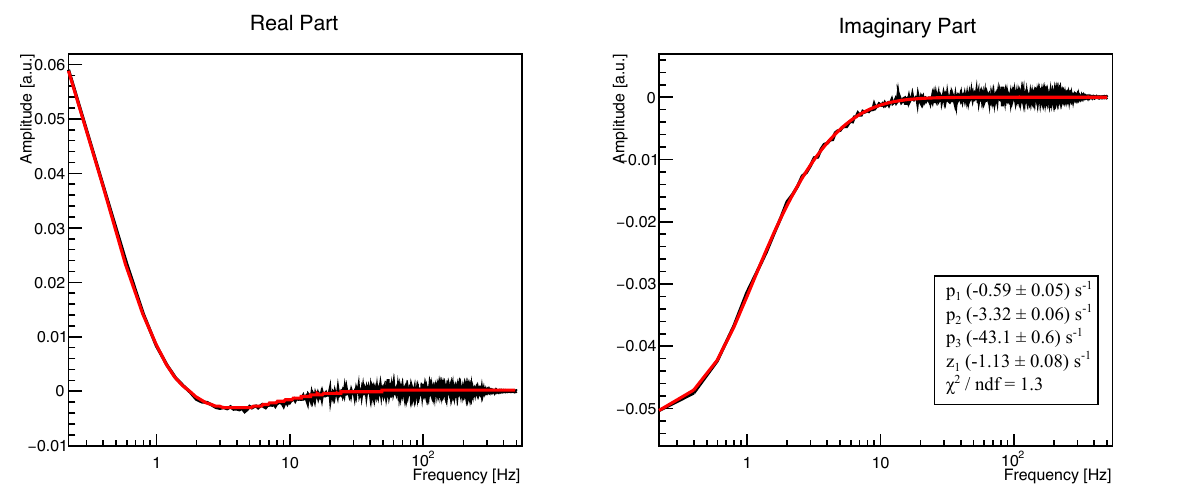}
\end{minipage}
\begin{minipage}[b]{\linewidth}
 \centering
\includegraphics[scale=0.5]{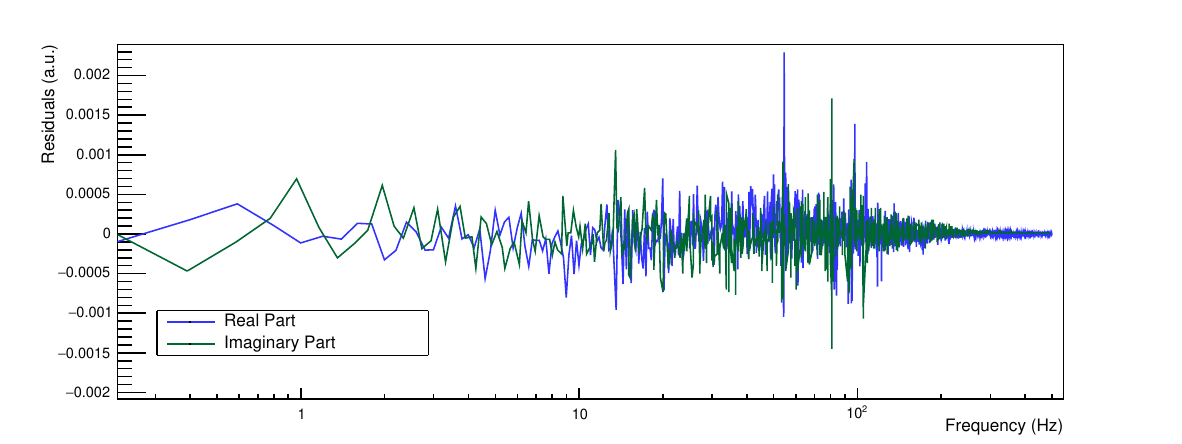}\\
\end{minipage}\\
\caption{[Top] Example: Discrete Fourier transform of the pulse from Fig.\ref{fig:example3p1z_TeO_Q0}. The pulse real and imaginary parts are fit in the frequency domain: the best fit corresponds to the red overlaid curves). [Bottom] Fit residuals for both real and imaginary parts.
}
\label{fig:example3p1z_TeO_Q0_freq}
\end{figure}

\subsection{Template with four real poles and one zero, and effect of the electro-thermal feedback}

Pulses from CUORE \cite{Fiorini:1998gj,Alduino:2017ehq,Adams:2019jhp,Nutini:2020vtd} detectors appear to have a slightly more complex shape than the CUORE-0 ones, discussed in the previous section, even if the detectors are almost identical. The peculiarities of the  CUORE detectors with respect to the CUORE-0 ones lie in the NTD type and the lower operating temperature.

\begin{figure}[htbp]
\centering
\includegraphics[width=0.9\textwidth]{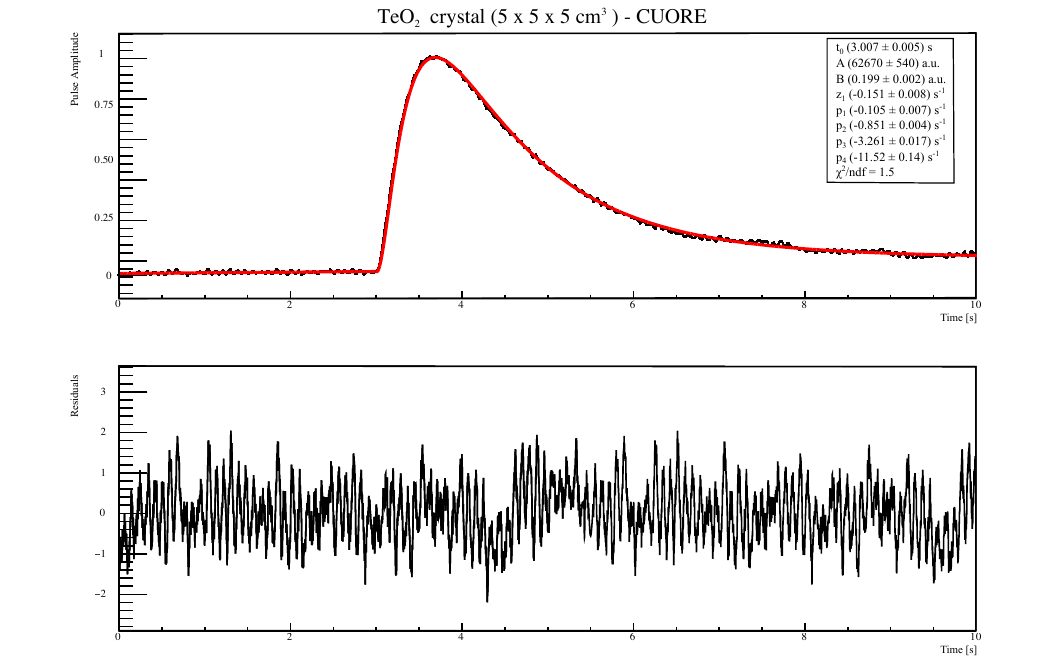}
\caption{Example: Pulse from a TeO$_2$ macro-calorimeter read by a NTD operated at low bias current (from CUORE experiment). The pulse is fit with a signal template with 4 real poles ($p_1$,$p_2$,$p_3$,$p_4$) and 1 zero ($z_1$).}
\label{fig:example4pRe1z}
\end{figure}

In case of low bias voltages applied to the NTDs, we see that the CUORE-0 like description does not hold. The pulse rise time is still faster than the two decay times, however the rising edge of the pulse appears not a pure exponential and slightly smoothens at the bottom; another pole (a faster decay constant $p_4$) is then necessary to describe the pulses. Four real poles and one zero are then defined: $ p_4 < p_3 < p_2 < z_1 < p_1 < 0 $, such that $p_3$ is related to the rise time constant and ($p_1$,$p_2$,$p_4$) to the decay ones (see Fig.\ref{fig:example4pRe1z}). We build then a transfer function $H(s)$ with 4 poles and 1 zero:

\begin{align}\label{eq:4poles}
H(s) = \frac{(s-z_1)}{(s - p_1)  (s - p_2) (s - p_3) (s - p_4)} 
\end{align}

\begin{figure}[htbp]
\centering
\includegraphics[width=0.9\textwidth]{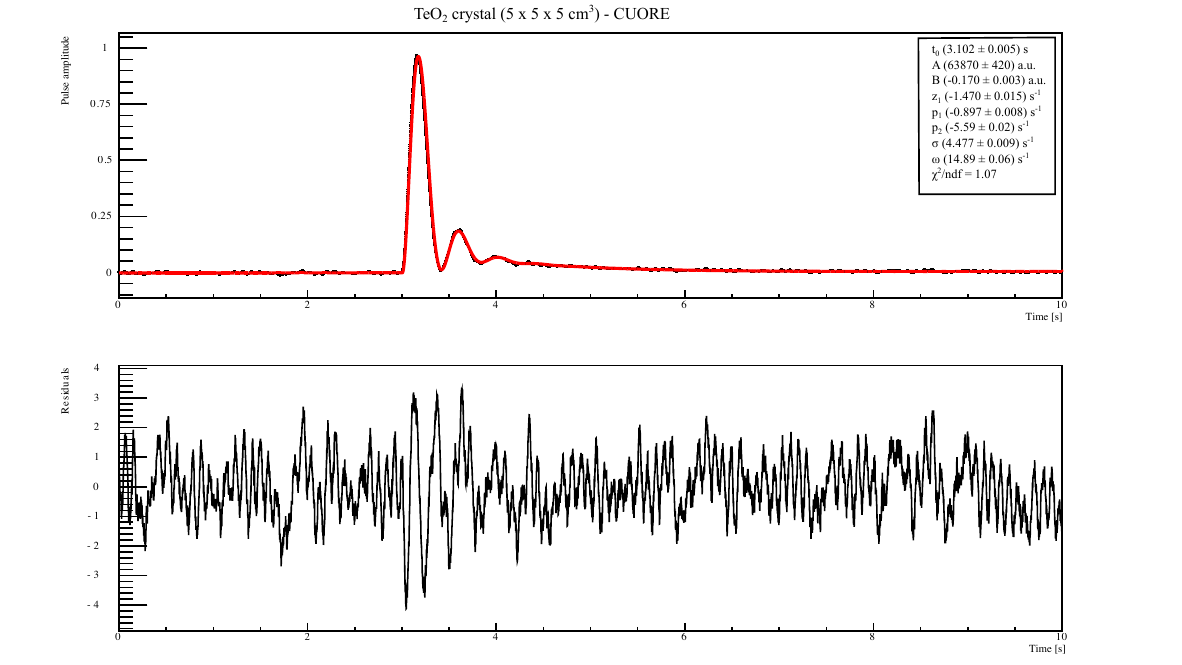}
\caption{Example: Pulse from a TeO$_2$ macro-calorimeter read by a NTD operated at high bias current (from CUORE experiment). This condition of operation makes the pulse showing a damped oscillation on its falling edge. The pulse is fit with a signal template with 2 real poles ($p_1$,$p_2$), 2 complex conjugate poles ($p_3$=$p_4^*$) and 1 zero ($z_1$). 
}
\label{fig:example2pRe2pcc1z}
\end{figure}

The NTD polarization introduces a feedback due to the Joule heating induced by the current of the biasing circuit ({\it{electro-thermal feedback}}). We can treat this effect as a feedback applied to the response function of the thermal circuit which modifies the poles position. It can happen that two poles collapse to the same value of the real part. By that moment, the two poles become a complex conjugate pair (with negative real part) \cite{Alfonso:2020yee, Nutini:2019jzm} : 
$$ p_3 = p_4^{\star}  \text{ :} \qquad p_3 = \sigma + i \omega  \text{ , } p_4 = \sigma - i \omega \text{ ,} \qquad \sigma < 0 $$
The presence of complex conjugate poles in the detector response functions leads to the appearance of an oscillatory term which tends to deform the falling edge of the pulse. \\
The detector response function $h(t)$ is then built as:
\begin{align}\label{eq:qPulse}
h(t) = B + A \cdot [h_1(t-t_0)+ h_2(t-t_0)]\\
h_1(t)= r_1 e^{p_1 (t-t_0)} + r_2 e^{p_2(t-t_0)}\\
h_2(t) = r_3 e^{\sigma (t-t_0)} cos(\omega(t-t_0) + atan(\phi))\\
\phi = -i \frac{(p_3 -z_1)}{(p_3-p_1)(p_3-p_2)(2 \omega)}
\end{align}
where $h_1(t)$ corresponds to the real poles ($p_1$, $p_2$) contribution, $h_2(t)$ to the complex conjugate poles ($p_3$, $p_4$). \\
For increasing bias and when the NTD operates close the I-V curve inversion point, the imaginary part $\omega$ becomes larger, in absolute value, than $\sigma$; 
in this case, an evident damped oscillation appears on the falling edge of the pulse. 
An example of the detector response function with 2 real negative poles, 2 complex conjugate poles and 1 zero is reported in Fig.\ref{fig:example2pRe2pcc1z}, where $p_2 < \sigma < z_1 < p_1 < 0$. We can see how the position of the complex conjugate pair of poles ($p_3,p_4$) between the two real poles, affects the pulse shape in the time-domain; a slow damped oscillation is superimposed to the exponential falling edge of the pulse.

\section{Conclusions}
An effective approach allowing to build analytical templates that well describe the pulse shape of the thermal signals for calorimeters coupled with thermistors was introduced. 
The analytical functions described in the previous sections appear to be able to reproduce with good accuracy thermal pulses of various detectors independently from the different physical quantities which contributes to the signal generation.

\section*{Acknowledgments}

The authors are grateful to the CUORE collaborators and the colleagues from the Milano Bicocca group of cryogenics and rare events for the useful and productive discussions and for the possibility to test the proposed pulse analysis approach on multiple sets of data from low temperature detectors.

\bibliographystyle{elsarticle-num}
\bibliography{main}

\end{document}